\newcommand*{\ATLASLATEXPATH}{./}

\documentclass[cernpreprint,texlive=2014,txfonts,USenglish]{\ATLASLATEXPATH atlasdoc}

\usepackage[subfigure=true]{\ATLASLATEXPATH atlaspackage}
\usepackage{\ATLASLATEXPATH atlasbiblatex}

\usepackage{\ATLASLATEXPATH atlascontribute}

\usepackage{\ATLASLATEXPATH atlasphysics}

\graphicspath{{logos/}{figures/}}

\usepackage{AHTottbar-defs}

\usepackage{multirow}

\newcommand\mttrec{\ensuremath{m_{t\bar{t}}^{\mathrm{reco}}}}

\def\met{\ensuremath{E_{\mathrm{T}}^{\mathrm{miss}}}}

\def\tanb {\ensuremath{\tan\beta}}
\def\ttbar {\ensuremath{t\bar{t}}}

\def\pt{\ensuremath{p_{\mathrm{T}}}}

\def\mtw{\ensuremath{m_{\mathrm{T}}^W}}

\def\ifb{\ensuremath{\mathrm{fb}^{-1}}}

\def\pt{\ensuremath{p_{\mathrm{T}}}}

\usepackage{verbatim}

\usepackage{siunitx}
\hypersetup{pdftitle={ATLAS draft},pdfauthor={The ATLAS Collaboration}}
\AtlasTitle{
Search for heavy Higgs bosons $\boldsymbol{A/H}$ decaying to a top quark pair in $\boldsymbol{pp}$ collisions at $\boldsymbol{\sqrt{s} = 8}$~TeV with the ATLAS detector 
}

\author{ATLAS Collaboration}

\AtlasRefCode{EXOT-2016-04}

\PreprintIdNumber{CERN-EP-2017-134}

\AtlasJournal{Phys.\ Rev.\ Lett.}
\AtlasJournalRef{PRL 119, 191803 (2017)}
\AtlasDOI{10.1103/PhysRevLett.119.191803}

\AtlasAbstract{%
A search for heavy pseudoscalar ($A$) and scalar ($H$) Higgs bosons decaying into a top quark pair ($t\bar{t}$) has been performed with 20.3~fb$^{-1}$
of proton--proton collision data collected by the ATLAS experiment at the Large Hadron Collider at a center-of-mass energy $\sqrt{s} = 8$~TeV.
Interference effects between the signal process and Standard Model $t\bar{t}$ production, 
which are expected to distort the signal shape from a single peak to a peak--dip structure, are taken into account.
No significant deviation from the Standard Model prediction is observed in the $t\bar{t}$ invariant mass spectrum in final states with an electron or muon, 
large missing transverse momentum, and at least four jets. 
The results are interpreted within the context of a type-II two-Higgs-doublet model.
Exclusion limits on the signal strength are derived as a function of the mass $m_{A/H}$ and the ratio
of the vacuum expectation values of the two Higgs fields, $\tan\beta$, for $m_{A/H} > 500$~GeV. 
}

\begin{document}

\maketitle





\section{Introduction}

The production of new particles at the Large Hadron Collider (LHC) with masses close to the TeV scale 
is predicted by many models of physics beyond the Standard Model (SM).
In this Letter, a search for massive pseudoscalar and scalar resonances 
decaying into a top--antitop quark pair ($t\bar{t}$) is presented. It is the first search in this final state to take into account the significant interference between 
the signal and the background from SM $t\bar{t}$ production.
The search is conducted on a sample of $pp$ collision data with an integrated luminosity of 20.3 fb$^{-1}$ at a center-of-mass energy $\sqrt{s}$ = 8~TeV,
collected with the ATLAS detector~\cite{Aad:2008zzm}. 

New pseudoscalar ($A$) and scalar ($H$) states coupling strongly to $t\bar{t}$ are predicted by a class of models in which the Higgs sector is extended 
to include a second Higgs doublet, the two-Higgs-doublet models (2HDMs)~\cite{HDMpaper}. These models are motivated by many theories beyond the SM,
such as supersymmetry~\cite{Golfand:1971iw,VOLKOV1973109,WESS197439,WESS197440,FERRARA1974413,SALAM1974353} and axion models~\cite{KIM19871}.
In 2HDMs of type II~\cite{HDMpaper}, such as the Minimal Supersymmetric Standard Model (MSSM)~\cite{FAYET1976159,FAYET1977489,FARRAR1978575,FAYET1979416,DIMOPOULOS1981150}, 
these states decay predominantly into $t\bar{t}$ pairs
if $m_{A/H}\geq 500$~GeV and the ratio of the vacuum expectation values of the two Higgs fields, $\tanb$, is small ($\tanb \lesssim 3$). 

To date, this parameter region has not been probed directly by searches in other final states~\cite{Aad:2015pla,HIGG-2015-08,CMS-HIG-13-025,CMS-HIG-14-017,CMS-HIG-14-034,Djouadi:2015jea}
or by previous searches for $t\bar{t}$ resonances~\cite{CMS-B2G-16-015,Paper8TeV,CMS-B2G-13-001,PhysRevLett.110.121802,PhysRevD.85.051101}.
The latter, which aim to identify resonant excesses in the $t\bar{t}$ invariant mass ($m_{t\bar{t}}$) spectrum, 
have a reduced sensitivity to 2HDM signatures as they do not take into account interference effects between the signal and the dominant background from SM $t\bar{t}$ production. 
These are significant for (pseudo)scalar Higgs bosons with masses above the $t\bar{t}$ production threshold where the interference between the gluon--gluon ($gg$) initiated loop production
and the irreducible background from SM $t\bar{t}$ production yields a non-negligible imaginary term in the amplitude,
which at the LHC is dominated by $gg \rightarrow t\bar{t}$ production~\cite{1984PhLB,Dicus:1994bm,Bernreuther:1997gs,Frederix:2007gi,Carena:2016npr,BuarqueFranzosi:2017jrj}. 
As a result of the interference, the signal shape is distorted from a Breit--Wigner peak to a peak--dip structure. 

The results of the search are interpreted in a CP-conserving type-II 2HDM with a softly broken $Z_2$ symmetry~\cite{Gunion:2002zf}.
The lighter of the two neutral CP-even states, $h$, is assumed to be the Higgs boson discovered at a mass of $m_h = 125$~GeV~\cite{HIGG-2012-27,CMS-HIG-12-028}
with couplings as predicted by the SM. This corresponds to the condition $\sin{(\alpha - \beta)} = 1$, referred to as the \textit{alignment limit},
where $\alpha$ denotes the mixing angle between the two CP-even states.
The parameter $m_{12}$ of the $Z_2$ breaking term of the potential is taken to be $m_{12}^2 = m_A^2 \tanb/(1+\tan^2\beta)$.
In this model, the production cross-sections and widths of $A$ and $H$, as well as the signal shape, are determined by $\tanb$ and the masses $m_A$ and $m_H$.
The search results are derived assuming mass degeneracy, $m_H = m_A$, such that both processes contribute to the $m_{t\bar{t}}$ spectrum,
a scenario motivated, for example, by the MSSM~\cite{Gunion:2002zf}. We also consider two scenarios in which only the interference pattern of either $A$ or $H$
appears in the $m_{t\bar{t}}$ spectrum~\cite{Footnote1}.

\section{Data and Monte Carlo samples}
This analysis closely follows the resolved-topology analysis in Ref.~\cite{Paper8TeV}. 
Events with signatures compatible with $t\bar{t}\rightarrow W^+bW^-\bar{b}$, with one $W$ boson decaying hadronically and the other leptonically, 
the lepton-plus-jets channel ($\ell$+jets, $\ell = e, \mu$),
were collected using single-electron and single-muon triggers.
The trigger efficiency is constant in the transverse momentum ($p_T$) of leptons with $\pt > 25$~GeV~\cite{PERF-2013-03,TRIG-2012-03}.
The dominant background arises from SM $t\bar{t}$ production, followed by a contribution from $W$+jets processes.
Data-driven techniques were used to normalize the $W$+jets background contribution
and to estimate the background from multijet events.
All other background processes were estimated using Monte Carlo (MC) simulation. 
The background estimates for all processes are identical to those in Ref.~\cite{Paper8TeV}. 

\FloatBarrier

The signal process $gg \rightarrow A/H \rightarrow t\bar{t}$, including the decays of the top quarks and resulting $W$ bosons,
was simulated using \MGMCatNLO~\cite{Alwall:2014hca}~v2.3.3 with the model of Ref.~\cite{MGmodel},
which implements the $A/H$ production through loop-induced gluon--gluon fusion
with loop contributions from top and bottom quarks at leading order (LO) in QCD.
The CT10 set~\cite{Lai:2010vv} of parton distribution functions (PDFs) was used and the renormalization and factorization scales were set to $\sqrt{\sum_{\textrm{decay products}} (\pt^2+m^2)}$.

\begin{figure}
  \centering
    \includegraphics[width=0.48\textwidth]{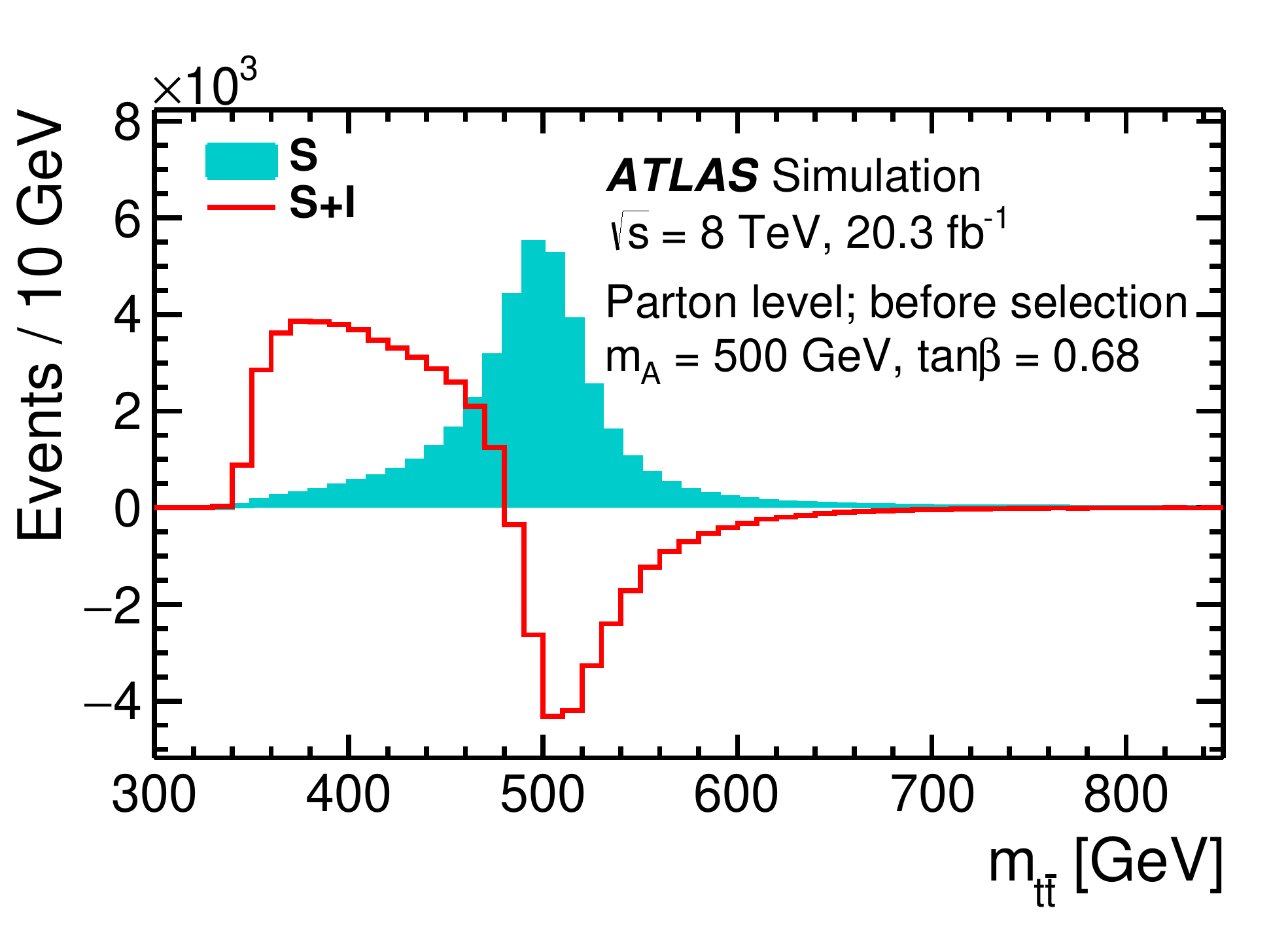}
  \caption{
    \label{fig:ParticleLevel}
    	Distributions of the invariant mass of the \ttbar~pair from the decay of a pseudoscalar $A$ of mass $m_A=500$~GeV 
	before the emission of final-state radiation and before the parton shower
	for the pure resonance $S$ (filled) and signal+interference contribution $S+I$ (unfilled).
	Events from all \ttbar~decay modes are included.
  }
\end{figure}

For the statistical interpretation, the $t\bar{t}$ invariant mass distributions in the signal regions in data were compared to a combination of the expected distributions from all background processes $B$,
the pure signal process $S$, and the signal-plus-interference component $S+I$ for a given signal hypothesis, as illustrated in Eq.~(\ref{eq:S+I+B}) below.
The most reliable description of the \ttbar~background~\cite{ATL-PHYS-PUB-2015-002}
is obtained at next-to-leading order (NLO) with {\textsc Powheg-Box}~\cite{Frixione:2007nw,Nason:2004rx,Frixione:2007vw,Alioli:2010xd} + {\textsc Pythia6}~\cite{Sjostrand:2006za}.
Therefore, the $S+I$ contribution was modeled separately from this background process
by modifying the \MGMCatNLO~software to remove the pure SM \ttbar~process to yield only the $S+I$ contribution 
on an event-by-event basis. The nominal \ttbar~background prediction in $m_{t\bar{t}}$ is in good agreement with that obtained from~\MGMCatNLO~in all signal regions.
The $S+I$ events obtained with the modified software can have positive or negative weights.
Figure~\ref{fig:ParticleLevel} shows the \ttbar~invariant mass distributions for the $S$ and $S+I$ components in a model with $\tanb=0.68$ and a pseudoscalar of mass $m_A=500$~GeV.
The $S+I$ component exhibits a peak-dip structure with the minimum around $m_{A/H}$ for all signal hypotheses studied in this search. The width of both the $S$ and $S+I$
distribution decreases with increasing $\tanb$.

The $S+I$ distributions from the modified \MGMCatNLO~software were validated against those from the unmodified program. The latter were obtained by generating
a large inclusive sample $S+I+B_{t\bar{t}}$ for a given parameter point and a LO SM \ttbar~background $B_{t\bar{t}}$ sample with the same generator settings.
The difference between the resulting two $m_{t\bar{t}}$ distributions corresponds to the $S+I$ component, which agrees with that obtained with the modified software
within 0.4\% across the whole spectrum. The difference is taken as a systematic uncertainty in $S+I$.

{\textsc Pythia6} with the Perugia 2011c set of tuned parameters~\cite{Skands:2010ak} was used to model the parton shower and hadronization for all $S$ and $S+I$ samples
and the stable particles obtained after hadronization were passed through the ATLAS fast detector simulation~\cite{ATL-PHYS-PUB-2010-013}.
The effects of additional collisions within the same or nearby bunch crossings were simulated by overlaying additional $pp$ collisions, 
simulated with {\textsc Pythia} v8.1~\cite{Sjostrand:2007gs}, on each event. 
Correction factors were applied to adjust the trigger and selection efficiencies in simulated events to those measured in data.
The $S$ and $S+I$ samples with this setup were generated separately for pseudoscalar and scalar Higgs bosons.

Event samples for both the $S$ and $S+I$ components for different values of $(m_{A/H},\tanb)$ were obtained from signal samples $S$ after the detector simulation by applying an event-by-event reweighting.
This reweighting substantially reduces the computing time required.
The weight is the ratio of the \MGMCatNLO~matrix elements, 
calculated from the four-momenta of the incoming gluons and outgoing top quarks of the generated event with the new and the old values of $(m_{A/H},\tanb)$, respectively.
All $S+I$ and a small number of $S$ samples were obtained through reweighting.
Signal hypotheses with $m_{A/H}<500$~GeV were not considered as they require an accurate modeling of the Higgs boson decay into virtual top quarks and the implementation of higher-order corrections
that are not available in the \MGMCatNLO~model.
The requirement $\tanb\geq 0.4$ was imposed to ensure the perturbativity of the top-quark Yukawa coupling~\cite{HDMpaper}.

Correction factors $K_S$ were applied to normalize the generated signal ($S$) cross-section to the value calculated
at partial next-to-next-to-leading-order (NNLO) precision in QCD~\cite{Eriksson:2009ws,sushi1,sushi2}. The correction factor for the interference component $I$ is $K_I=\sqrt{K_S\times K_B}$,
as suggested in Ref.~\cite{Hespel:2016qaf}, where $K_B=1.87$ is the correction factor to normalize the total cross-section of the SM $t\bar{t}$ background generated at LO with {\textsc MadGraph}
to the cross-section calculated at NNLO accuracy in the strong coupling constant \alphas, including resummation of 
next-to-next-to-leading-logarithmic soft gluon terms. The values of $K_S$ range between two and three for the tested signal hypotheses.

\section{Event selection}
The event selection criteria for the signal regions provide a high selection efficiency for $t\bar{t}$ events.
Only events with a \textit{resolved topology}, in which the three jets
from the hadronically decaying top quark are well separated in the detector, are selected.
This is the most efficient selection strategy for signal hypotheses with $m_{A/H} < 800$~GeV.
Events with a \textit{merged topology}, in which the top quark is reconstructed as a single jet,
are not considered.
The event reconstruction and selection criteria are identical to those in Ref.~\cite{Paper8TeV}
except that events that would satisfy the criteria for both topologies are classified as ``resolved'' instead of ``merged''.

Events are required to contain exactly one isolated electron~\cite{PERF-2016-01} or muon~\cite{PERF-2014-05} with $\pt > 25$~GeV and pseudorapidity $|\eta| < 2.5$~\cite{Footnote}.
Events must have large missing transverse momentum, $\met > 20$~GeV, computed as the magnitude of the negative vector sum of lepton and jet transverse momenta~\cite{PERF-2014-04}.
In addition, $\met + \mtw > 60$~GeV, is required to further suppress the contribution from multijet events, where $\mtw$ is the lepton--$\met$ transverse mass~\cite{Paper8TeV}.
Events must contain at least four hadronic jets with $\pT>25$~GeV and $|\eta|<2.5$, reconstructed using the anti-$k_t$ algorithm~\cite{AKT:2008,Cacciari:2011ma} 
with radius parameter $R=0.4$.
Jets from additional collisions in the same bunch crossing are rejected using dedicated tracking and vertex requirements~\cite{PERF-2014-03}.
At least one of the jets must be identified as originating from the decay of a $b$-hadron ($b$-jet) using a multivariate
tagging algorithm with a 70\% efficiency for $b$-jets and light-quark and gluon mistag rates of 0.5-2\%~\cite{PERF-2012-04}.


\section{Event reconstruction}

Jets are assigned to the top quarks using a $\chi^2$ algorithm that relies on kinematic constraints
and the expected values of the top quark and $W$ boson masses~\cite{Paper8TeV}.
The invariant mass $m_{t\bar{t}}^{\textrm{reco}}$ of the candidate \ttbar~pair is reconstructed 
from the four selected jets, the lepton, and the $\met$ vector.
The experimental resolution for the $t\bar{t}$ invariant mass is 8\% for $m_{A/H}=500$~GeV.
Events in the $e$+jets and $\mu$+jets channels are classified into three categories,
based on whether a $b$-tagged jet was assigned to either the hadronically or the semileptonically decaying top quark, or to both of them.
Each category defines a signal region, hence six orthogonal signal regions are used in the statistical analysis.


\section{Systematic uncertainties}

\begin{table}[tbh]
\begin{center}
\caption[]{
Average impact of the dominant uncertainties on the estimated yields for the total background and for a pseudoscalar $A$
with $m_A = 500$~GeV and $\tanb=0.68$ in percent of the nominal value for all signal regions combined.
Only uncertainties with a yield impact $>0.5\%$ are shown.
A bar ($-$) indicates that an uncertainty is not applicable to a sample.
\label{tab:systimpact}
}
\begin{tabular}{l|r|r|r}
\hline
\hline
Systematic uncertainties [\%]  & Total bkg & $S$  & $S+I$ \hspace{1ex}\\
\hline
\hline
Luminosity~\cite{DAPR-2013-01}          				&  $1.7$     &  $1.9$        &  $1.9$\\ \hline                 
PDF               							&  $2.5$     &  $2.1$        &  $12$\\ \hline 
$t\bar{t}$ initial-/final-state radiation               		&  $3.2$     &  $-$          &  $-$\\ \hline 
$t\bar{t}$ parton shower + fragmentation   				&  $4.9$     &  $-$          &  $-$\\ \hline     
$t\bar{t}$ normalization            	&  $5.7$     &  $-$          &  $-$\\ \hline 
$t\bar{t}$ event generator     		&  $0.5$     &  $-$          &  $-$\\ \hline 
Top quark mass           		&  $0.5$     &  $2.2$        &  $13$\\ \hline   
Jet energy scale       			&  $6.4$     &  $4.9$        &  $9.3$ \\ \hline 
Jet energy resolution   		&  $1.3$     &  $1.6$        &  $1.7$ \\ \hline 
$b$-tagging: $b$-jet efficiency     	&  $1.5$     &  $1.3$        &  $1.1$\\ \hline
$b$-tagging: $c$-jet efficiency     	&  $0.2$     &  $0.2$        &  $0.8$\\ \hline 
Electron efficiency              	&  $0.3$     &  $0.4$        &  $0.7$\\ \hline 
Muon efficiency                  	&  $0.9$     &  $1.0$        &  $1.0$\\ \hline 
Signal MC scales	                &  $-$       &  $7.3$        &  $7.3$\\ \hline
Reweighting		                &  $-$       &  $-$         &  $5.0$\\ \hline
MC statistical uncertainty         	&  $0.5$     &  $2.4$        &  $11$\\ \hline         
\hline                                                                                                       
Total uncertainty       			&  $11$    &  $10$        &  $25$\\ \hline 
\hline
\end{tabular}
\end{center}
\end{table} 

The impact of the systematic uncertainties on both the normalization and the shape of the $m_{t\bar{t}}^{\textrm{reco}}$ distributions is taken into account. 
The average impact of the dominant uncertainties on the event yields is summarized in Table~\ref{tab:systimpact}.

The experimental uncertainties with the largest impact on the event yields and the shape of the $m_{t\bar{t}}^{\textrm{reco}}$ distributions
are those related to the jet energy scale (JES) and the jet energy resolution (JER)~\cite{PERF-2011-03,PERF-2012-01},
followed by uncertainties on the $b$-tagging efficiency and misidentification rates~\cite{PERF-2012-04}.
The uncertainties related to leptons include those in the reconstruction and isolation efficiency, the single-lepton triggers,
and the energy scale and resolution~\cite{PERF-2016-01,PERF-2014-05}. 

The uncertainty of 6.5\% in the NNLO+NNLL cross-section for SM $t\bar{t}$ production, is the dominant uncertainty in the total background normalization~\cite{Paper8TeV}.
Modeling uncertainties affecting the shape of the $m_{t\bar{t}}^{\textrm{reco}}$ distribution
for the SM $t\bar{t}$ background are also taken into account.
These include uncertainties related to the choice of NLO event generator, the modeling of the parton shower and fragmentation,
the modeling of gluon initial- and final-state radiation, and the value of the top quark mass $m_{t}$.
Other sources of uncertainty related to the various background components are described in Ref.~\cite{Paper8TeV}.

The largest uncertainty in the modeling of the $S+I$ and $S$ components is related to the $\pm 1.0$~GeV uncertainty of the value $m_{t}=172.5$~GeV~\cite{ATLAS:2014wva}.
Uncertainties related to the choice of PDF set and renormalization and factorization scales are also considered.
The latter is estimated by varying the scales by factors of 0.5 and 2.0, which yields a constant $\pm 7.3\%$ variation across the \mttrec~spectrum.
An asymmetric variation, for which the bins at the low and high ends of the \mttrec~spectrum are taken as anticorrelated~\cite{Lindert:2017olm}
is also considered to estimate the impact of the scale variations on the shape of the \mttrec~spectrum.
For the $S+I$ samples, an additional constant $\pm 5$\% uncertainty is included to cover the difference between reweighted and generated distributions.

\section{Results}
A breakdown of the observed and expected event yields in the $e$+jets and $\mu$+jets
channels and their total uncertainties is shown in Table~\ref{tab:event_yields}.
Good agreement is found between the observed number of events in data and the expected total number of background events.

\begin{table}[tbh!]
\begin{center}
\caption{
Number of events observed in data and expected  number of background events after the event selection, before the profile-likelihood fit to the full dataset.
The uncertainty in the background yields is derived by summing all uncertainties in quadrature. 
The ``other bkg'' component comprises single top quark, $t\bar{t}+W/Z$, $Z$+jets, diboson and multijet production.
\label{tab:event_yields}
}
  \centering
\begin{tabular}{lr@{$~\pm~$}lr@{$~\pm~$}l}
\hline\hline
Type          & \multicolumn{2}{c}{$e$+jets}  & \multicolumn{2}{c}{$\mu$+jets} \\  \hline \hline
$t\bar{t}$ & 95\,000 &  11\,000 & 93\,000 &  11\,000 \\
$W$+jets & 6600 &  2100 & 7200 &  2300 \\
Other bkg. & 11\,200 &  1400 & 6100 &  600 \\
Total & 112\,800 &  13\,000 & 106\,300 &  12\,000 \\
\hline
Data & \multicolumn{2}{c}{115\,785} & \multicolumn{2}{c}{110\,218} \\\hline
\hline
\end{tabular}
\end{center}
\end{table}


\begin{figure}
  \centering
    \includegraphics[width=0.48\textwidth]{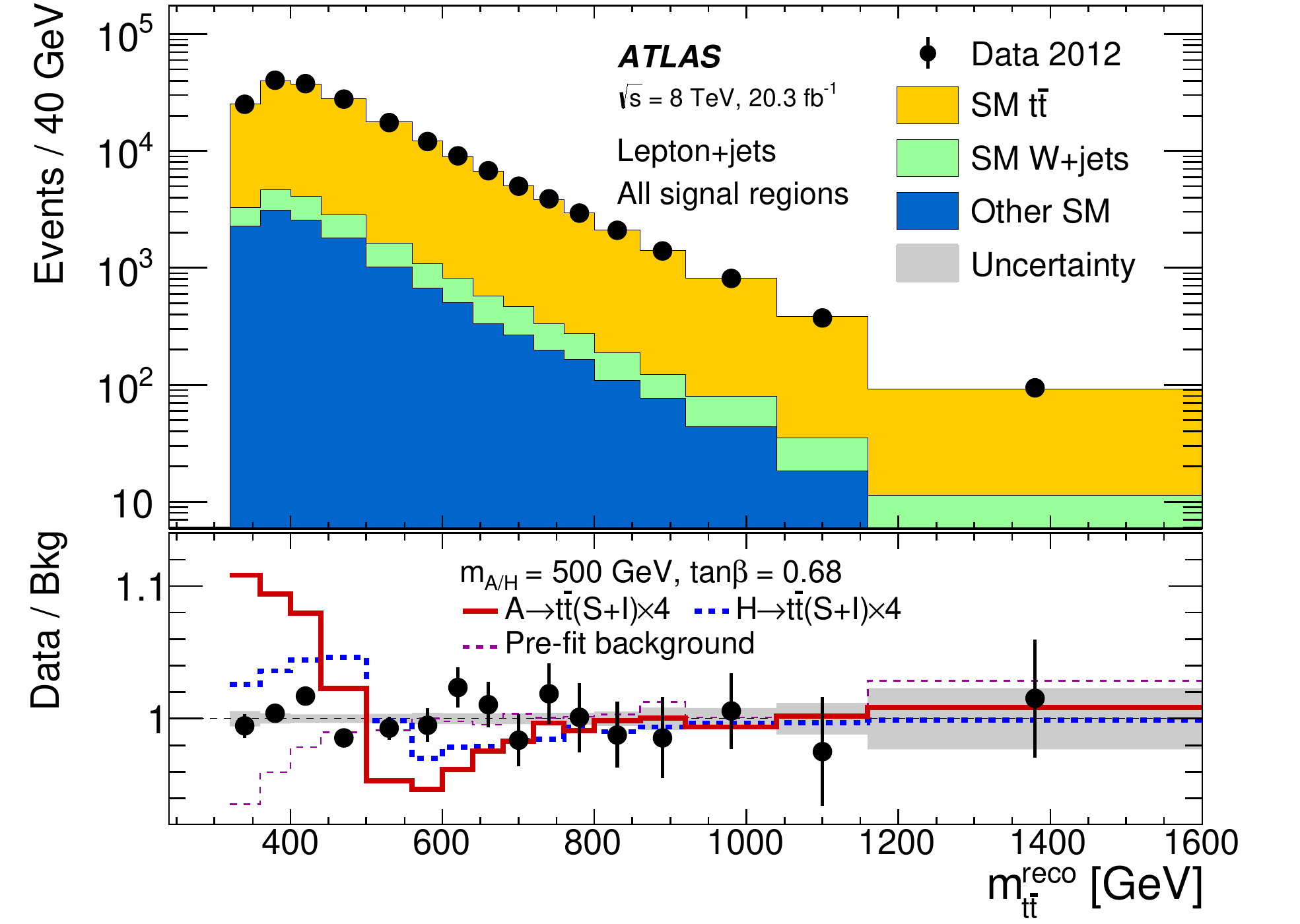}
  \caption{
    \label{fig:postFit} 
    Distribution of $m_{\textrm{\ttbar}}^{\textrm{reco}}$ for the data and the expected background after the profile-likelihood fit
    under the background-only hypothesis for all signal regions combined.
    The lines in the bottom panel show the individual $S+I$ distributions (scaled by a factor of four)
    for a pseudoscalar $A$ (solid) and scalar $H$ (bold dashed) with $m_{A/H}=500$~GeV and $\tan\beta=0.68$ relative to the total background.
  }
\end{figure}

\FloatBarrier

The exclusion limits are derived separately for each signal hypothesis
from a profile-likelihood fit~\cite{Cowan:2010js} of the expected $\mttrec$ distributions to the observed ones
simultaneously in all signal regions, taking the statistical and systematic uncertainties into account as nuisance parameters~\cite{Paper8TeV}.
Only bins with $m_{\textrm{\ttbar}}^{\textrm{reco}} > 320$~GeV are considered to avoid threshold effects not well described by the simulation.
The shape of the binned \mttrec~distributions is parameterized in terms of the signal strength $\mu$~\cite{1984PhLB,Dicus:1994bm}:
\begin{equation}\label{eq:S+I+B}
  \mu\cdot S + \sqrt{\mu}\cdot I + B = (\mu -\sqrt{\mu})\cdot S + \sqrt{\mu} \cdot (S+I) + B.
\end{equation}
The fitted variable is $\sqrt{\mu}$ and the case $\mu = 1$ ($\mu = 0$) corresponds to the type-II 2HDM in the alignment limit (the background-only hypothesis).
This approach relies on the assumption that, for a given signal hypothesis, 
the shape of the $t\bar{t}$ invariant mass distributions for $S$ and $S+I$  
in Eq.~(\ref{eq:S+I+B}) does not change with $\mu$.
The terms $S$ and $S+I$ on the right-hand side of Eq.~(\ref{eq:S+I+B}) correspond to the $\mttrec$ distributions obtained from the $S$ and $S+I$ samples, respectively,
while $B$ stands for the expected $\mttrec$ distribution of the total background. 

The level of agreement between the observed and expected mass spectra is quantified in a fit under the background-only hypothesis 
in which only the nuisance parameters are allowed to vary.
The observed $\mttrec$ spectra are compatible with the post-fit expected spectra 
within the (constrained) uncertainty bands (Figure~\ref{fig:postFit}). 

\begin{figure}[h!]
  \centering
    \includegraphics[width=0.99\textwidth]{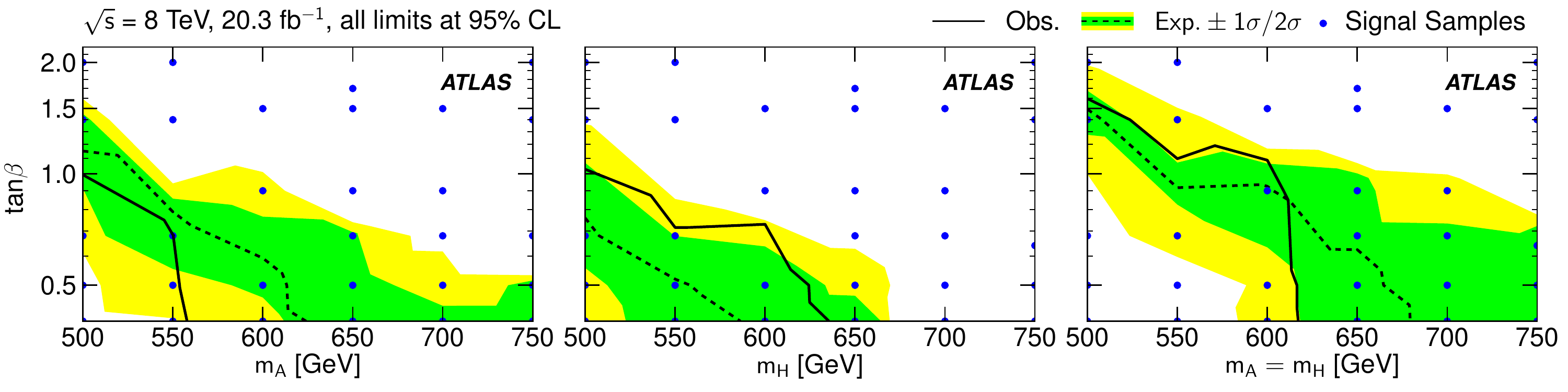}
  \caption{
    \label{fig:2DLimitsJamaica}
    The 95\% CL observed and expected exclusion regions for the type-II 2HDM ($\mu = 1$) considering only a pseudoscalar $A$ (left), 
    only a scalar $H$ (middle), and the mass-degenerate scenario $m_A = m_H$ (right).
    Blue points indicate parameter values at which signal samples are produced.
  }
\end{figure}

The upper limits on $\mu$ at 95\% confidence level (CL) are obtained with the CL$_{\textrm{s}}$ method~\cite{Read:2002hq}
for a number of $(m_{A/H},\tanb)$ values.
The upper limits at intermediate points are obtained from a linear interpolation among the three closest points.
In Figure~\ref{fig:2DLimitsJamaica}, the observed and expected exclusion regions for the type-II 2HDM ($\mu = 1$)
are shown for the three scenarios discussed in the introduction.
The excluded values of $\tanb$ for the different mass hypotheses are listed in Table~\ref{tab:excl}.

\begin{table}[h!]
\begin{center}
\caption{
The 95\% CL observed and expected exclusion limits on $\tanb$ for a type-II 2HDM in the alignment limit considering only a pseudoscalar $A$ (left), 
only a scalar $H$ (middle), and the mass-degenerate scenario $m_A = m_H$ (right). A bar (--) indicates that no value of $\tanb\geq0.4$ is excluded.
\label{tab:excl}
}
  \centering
  \begin{tabular}{llcccccccc}\hline\hline                                                                                        
      	Mass 	      && \multicolumn{2}{c}{$m_A$}     &  \hspace*{0.8cm}& \multicolumn{2}{c}{$m_H$}&  \hspace*{0.8cm}& \multicolumn{2}{c}{$m_A=m_H$} \\ 
      	 {[GeV]}      &\tanb: & obs. & exp. &&     obs. & exp. &&      obs. & exp.	\\
        \hline\hline
      500        && $<1.00$ 	& $<1.16$ 	&&      $<1.00$     	& $<0.77$ 	&&      $<1.55$ 	& $<1.50$	\\
      550        &&$<0.69$ 	& $<0.79$ 	&&      $<0.72$     	& $<0.52$ 	&&      $<1.10$ 	& $<0.92$	\\
      600        && -- 	& $<0.59$ 	&&      $<0.73$		& --		&&      $<1.09$		& $<0.93$	\\
      650        && -- 	& -- 		&&      --		& --		&&      -- 		& $<0.62$	\\
        \hline\hline                                  
\end{tabular}
\end{center}
\end{table}

\FloatBarrier

\section{Conclusion}
In conclusion, the search for massive pseudoscalar and scalar resonances decaying to $t\bar{t}$ in 
20.3~fb$^{-1}$ of $pp$ collisions at $8$~TeV recorded by the ATLAS experiment, 
yields no statistically signifiant deviations from the SM prediction.
The results are interpreted in a type-II 2HDM in the alignment limit,
and upper limits are set on the signal strength $\mu$ at 95\% CL 
in the $m_{A/H}$ versus $\tanb$ plane.
Unlike previous searches for $t\bar{t}$ resonances, this analysis 
takes into account interference effects between the signal process and the background from SM $t\bar{t}$ production. 
It tightens significantly the previously published constraints on the 2HDM parameter space
in the low \tanb\ and high mass ($m_{A/H} > 500$~GeV) region.


We thank CERN for the very successful operation of the LHC, as well as the
support staff from our institutions without whom ATLAS could not be
operated efficiently.

We acknowledge the support of ANPCyT, Argentina; YerPhI, Armenia; ARC, Australia; BMWFW and FWF, Austria; ANAS, Azerbaijan; SSTC, Belarus; 
CNPq and FAPESP, Brazil; NSERC, NRC and CFI, Canada; CERN; CONICYT, Chile; CAS, MOST and NSFC, China; COLCIENCIAS, Colombia; MSMT CR, MPO CR and VSC CR, 
Czech Republic; DNRF and DNSRC, Denmark; IN2P3-CNRS, CEA-DSM/IRFU, France; SRNSF, Georgia; BMBF, HGF, and MPG, Germany; GSRT, Greece; RGC, Hong Kong SAR, China; 
ISF, I-CORE and Benoziyo Center, Israel; INFN, Italy; MEXT and JSPS, Japan; CNRST, Morocco; NWO, Netherlands; RCN, Norway; MNiSW and NCN, Poland; FCT, Portugal;
MNE/IFA, Romania; MES of Russia and NRC KI, Russian Federation; JINR; MESTD, Serbia; MSSR, Slovakia; ARRS and MIZ\v{S}, Slovenia; DST/NRF, South Africa; MINECO, Spain; 
SRC and Wallenberg Foundation, Sweden; SERI, SNSF and Cantons of Bern and Geneva, Switzerland; MOST, Taiwan; TAEK, Turkey; STFC, United Kingdom; DOE and NSF, United States of America. 
In addition, individual groups and members have received support from BCKDF, the Canada Council, CANARIE, CRC, Compute Canada, FQRNT, and the Ontario Innovation Trust, Canada; 
EPLANET, ERC, ERDF, FP7, Horizon 2020 and Marie Sk{\l}odowska-Curie Actions, European Union; Investissements d'Avenir Labex and Idex, ANR, R{\'e}gion Auvergne and Fondation Partager le Savoir, France; 
DFG and AvH Foundation, Germany; Herakleitos, Thales and Aristeia programmes co-financed by EU-ESF and the Greek NSRF; BSF, GIF and Minerva, Israel; BRF, Norway; 
CERCA Programme Generalitat de Catalunya, Generalitat Valenciana, Spain; the Royal Society and Leverhulme Trust, United Kingdom.

The crucial computing support from all WLCG partners is acknowledged gratefully, in particular from CERN, the ATLAS Tier-1 facilities 
at TRIUMF (Canada), NDGF (Denmark, Norway, Sweden), CC-IN2P3 (France), KIT/GridKA (Germany), INFN-CNAF (Italy), NL-T1 (Netherlands), PIC (Spain), 
ASGC (Taiwan), RAL (UK) and BNL (USA), the Tier-2 facilities worldwide and large non-WLCG resource providers. 
Major contributors of computing resources are listed in Ref.~\cite{ATL-GEN-PUB-2016-002}.


\printbibliography

\onecolumn

\newpage 
\begin{flushleft}
{\Large The ATLAS Collaboration}

\bigskip

M.~Aaboud$^\textrm{\scriptsize 137d}$,
G.~Aad$^\textrm{\scriptsize 88}$,
B.~Abbott$^\textrm{\scriptsize 115}$,
O.~Abdinov$^\textrm{\scriptsize 12}$$^{,*}$,
B.~Abeloos$^\textrm{\scriptsize 119}$,
S.H.~Abidi$^\textrm{\scriptsize 161}$,
O.S.~AbouZeid$^\textrm{\scriptsize 139}$,
N.L.~Abraham$^\textrm{\scriptsize 151}$,
H.~Abramowicz$^\textrm{\scriptsize 155}$,
H.~Abreu$^\textrm{\scriptsize 154}$
% ...
\bigskip
\\
% ...
$^{12}$ Institute of Physics, Azerbaijan Academy of Sciences, Baku, Azerbaijan\\
% ...
$^{88}$ CPPM, Aix-Marseille Universit{\'e} and CNRS/IN2P3, Marseille, France\\
% ...
$^{115}$ Homer L. Dodge Department of Physics and Astronomy, University of Oklahoma, Norman OK, United States of America\\
% ...
$^{119}$ LAL, Univ. Paris-Sud, CNRS/IN2P3, Universit{\'e} Paris-Saclay, Orsay, France\\
% ...
$^{137}$ $^{(a)}$ Facult{\'e} des Sciences Ain Chock, R{\'e}seau Universitaire de Physique des Hautes Energies - Universit{\'e} Hassan II, Casablanca; $^{(b)}$ Centre National de l'Energie des Sciences Techniques Nucleaires, Rabat; $^{(c)}$ Facult{\'e} des Sciences Semlalia, Universit{\'e} Cadi Ayyad, LPHEA-Marrakech; $^{(d)}$ Facult{\'e} des Sciences, Universit{\'e} Mohamed Premier and LPTPM, Oujda; $^{(e)}$ Facult{\'e} des sciences, Universit{\'e} Mohammed V, Rabat, Morocco\\
% ...
$^{139}$ Santa Cruz Institute for Particle Physics, University of California Santa Cruz, Santa Cruz CA, United States of America\\
% ...
$^{151}$ Department of Physics and Astronomy, University of Sussex, Brighton, United Kingdom\\
% ...
$^{154}$ Department of Physics, Technion: Israel Institute of Technology, Haifa, Israel\\
$^{155}$ Raymond and Beverly Sackler School of Physics and Astronomy, Tel Aviv University, Tel Aviv, Israel\\
% ...
$^{161}$ Department of Physics, University of Toronto, Toronto ON, Canada\\
% ...
$^{a}$ Also at Department of Physics, King's College London, London, United Kingdom\\
$^{b}$ Also at Institute of Physics, Azerbaijan Academy of Sciences, Baku, Azerbaijan\\
$^{c}$ Also at Novosibirsk State University, Novosibirsk, Russia\\
$^{d}$ Also at TRIUMF, Vancouver BC, Canada\\
$^{e}$ Also at Department of Physics {\&} Astronomy, University of Louisville, Louisville, KY, United States of America\\
% ...
$^{*}$ Deceased
\end{flushleft}


\end{document}